\documentclass[%
 reprint,
 amsmath,amssymb,
 aps,
]{revtex4-2}

\usepackage{graphicx}
\usepackage{dcolumn}
\usepackage{bm}
\usepackage{amsmath}
\usepackage{breqn}
\usepackage[dvipsnames]{xcolor}
\usepackage{colortbl}
\usepackage[normalem]{ulem}

\definecolor{c1}{RGB}{255,242,242}
\definecolor{c2}{RGB}{255,198,198}
\definecolor{c3}{RGB}{255,142,142}
\definecolor{c4}{RGB}{193,193,255}
\definecolor{c5}{RGB}{255,255,250}
\definecolor{c6}{RGB}{255,177,177}
\definecolor{c7}{RGB}{142,142,255}
\definecolor{c8}{RGB}{212,212,255}
\definecolor{c9}{RGB}{245,245,255}

\begin{document}

\preprint{APS/123-QED}

\title{Cavity optomechanics in a fiber cavity: the role of stimulated Brillouin scattering}

\author{A. Beregi}%
\affiliation{Clarendon Laboratory$,$ University of Oxford$,$ Oxford OX1 3PU$,$ United Kingdom}%

\author{P. F. Barker}
 \email{p.barker@ucl.ac.uk}
\affiliation{Department of Physics and Astronomy$,$ University College London$,$ Gower Street$,$ London WC1E 6BT$,$ United Kingdom}

\author{A. Pontin}
\email{a.pontin@ucl.ac.uk}
\affiliation{Department of Physics and Astronomy$,$ University College London$,$ Gower Street$,$ London WC1E 6BT$,$ United Kingdom}

\begin{abstract}
\noindent We study the role of stimulated Brillouin scattering in a fiber cavity by numerical simulations and a simple theoretical model and find good agreement between experiment, simulation and theory. We also investigate an optomechanical system based on a fiber cavity in the presence on the nonlinear Brillouin scattering. Using simulation and theory, we show that this hybrid optomechanical system increases optomechanical damping for low mechanical resonance frequencies in the unresolved sideband regime. Furthermore, optimal damping occurs for blue detuning in stark contrast to standard optomechanics. We investigate whether this hybrid optomechanical system is capable cooling a mechanical oscillator to the quantum ground state.
\end{abstract}

\maketitle


\section{\label{sec:Intro}Introduction }

Cavity optomechanics explores the interaction between electromagnetic radiation in an optical cavity and a mechanical oscillator. The tremendous technological advancements over the last few decades have allowed the field to make an impact in both applications and foundational physics. Nowadays, many aspects of the quantum nature of the light-matter interaction have been observed, among these, ground state cooling \cite{Chan2011}, ponderomotive light squeezing~\cite{Brooks2012,regalsqueezing}, and quantum non-demolition measurements of field fluctuations~\cite{pontinQND}. More recently, levitation of nano- and micro-particles has been gaining more attention, since this platform has entered the quantum regime as well~\cite{Deli2020,novotnyalmostGS}, furthermore, levitation is considered one of the best candidate to study the quantum to classical transition and collapse models~\cite{Donadi2020,Pontin2020}.

Levitated optomechanical sensors offer ultrasensitive detection down to single molecule and single spin detection~\cite{singlemolecule,singlespin}, with application in acceleration sensing and gravimetry~\cite{mooreNanog,markusLia} and detection of weak forces~\cite{hendrik1,geracishort,geraciforce,Pontinforce}. A particularly intriguing application considers a $100$\,m long low finesse levitated optomechanical cavity for the detection of high frequency gravitational waves~\cite{PhysRevLett.110.071105} which could offer a strain sensitivity surpassing that of LIGO~\cite{advLIGO} and VIRGO~\cite{advVIRGO} in the $100$\,kHz frequency band. Later, it has been shown that a similar optomechanical experiment where the long free space cavity is substituted by an extrinsic fiber Fabry-Perot cavity can allow motional ground state cooling of a levitated microdisk, provided that nonlinear processes introduced by the fiber do not play a significant role~\cite{Pontin2018}. Among these, thermooptic noise, stimulated Brillouin scattering (SBS) and stimulated Raman scattering (SRS)~\cite{agrawal2006}.

In this paper, we investigate the effects of SBS in a fiber cavity optomechanical setup and show that in specific circumstances, one can exploit SBS to increase the optomechanical cooling rate. This system is a hybrid optomechanical system, since the optical cavity mode is coupled to the mechanical mode of the oscillator as well as to mechanical modes (lattice vibrations) in the optical fiber. Various hybrid optomechanical systems have already been studied. For example, it has been shown that if the cavity contained  two-level systems, enhanced cooling of the mechanical oscillator can be achieved~\cite{Restrepo2014} and that  bi- and tripartite entanglement can be created~\cite{Genes2008}. A similar enhancement has been show also when the two-level systems are embedded in the mechanical oscillator~\cite{Dantan2014}. Furthermore, a  cavity with an externally pumped gain medium as a two-level system was explored and it was shown that seeding this cavity externally results in cooling its own mirrors~\cite{Ge2013}. The system we consider here is fundamentally different from the two-level systems previously considered in optomechanics both from the microscopic point of view and the mechanisms of depletion. If two-level systems are between the cavity mirrors, the two relevant processes are absorption and stimulated emission of photons, both resulting in a change in photon number. In the case of SBS, a pump photon is absorbed while a phonon and a Stokes photon are emitted in the opposite direction of the pump photon propagation, conserving the photon number. Also, in the pumped gain medium case, the effects depend on the number of two-level systems in the excited state and, if the pump is sufficiently strong, the number of photons grow exponentially with propagation distance in the gain medium. In the case of an optically active medium, the depletion mechanism is the high number of photons in the cavity (at a constant pumping rate) opposed to the low number of pump photons in the case of SBS. Lastly, if two-level systems are present between the cavity mirrors, the detuning between cavity and atomic resonance plays an important role, while stimulated Brillouin scattering depends on pump frequency weakly (non-resonantly).
The hybrid fiber cavity optomechanical system considered here is capable of enhancing the optomechanical cooling in low-finesse, nearly sideband-resolved systems in general.  This effect is more robust, allowing a wider range of experimental parameters in contrast with the two-level systems between the cavity mirrors, where cooling depends strongly on atom-cavity and pump-atom detunings.

This paper is structured as follows. In Section~\ref{sec:Theory} we present a set of coupled differential equations that describe the evolution of the optical fields in a fiber cavity, which we solve numerically later. Moreover, we present a set of equations of motion using the formalism of quantum optics, which can be solved analytically when linearised around a steady-state solution. In Section~\ref{sec: Simulation vs experiment} we compare the predictions of the simulation and the linearised theory with experimental observations. We explore the possibility of increasing the optomechanical damping rate in Section~\ref{sec: enhanced cooling} and investigate the effects of various parameters. Finally, in Section~\ref{sec: Ground state cooling} we investigate whether the increased damping rate results in ground state cooling.

\section{\label{sec:Theory}Theory, simulation }

Differential equations describing the evolution of fields in a fiber cavity with nonlinear effects were derived by Ogusu~\cite{Ogusu2003}. These equations consider SBS with up to second order Stokes fields and first order anti-Stokes fields, self-phase modulation (SPM), cross-phase modulation (XPM) and four-wave mixing (FWM). Including all these effects in the theoretical model leads to complicated nonlinear differential equations. In our model, however, we put emphasis on SBS with first order Stokes fields only. The results in Ref.~\cite{Ogusu2003} demonstrate that second order Stokes and anti-Stokes scattering has a significantly higher threshold power than first-order Stokes scattering therefore we neglect these. This approximation also means that we can neglect FWM, since the terms describing these in the differential equations are proportional to either the anti-Stokes or second order Stokes amplitudes. Moreover, phase modulation effects can be neglected as the phase shift per unit optical power is approximately $0.1$~rad/W while the typical optical powers considered here are in the $30-80$/\,mW range. Therefore, the equations we simulate are
\begin{eqnarray}
\label{eqn:fieldeq1}
\frac{\partial E_p^+}{\partial z } = -\frac{\alpha}{2}E_p^+ - c_B \vert E_S^- \vert ^2E_p^+\\
 \frac{\partial E_p^-}{\partial z } = +\frac{\alpha}{2}E_p^- + c_B \vert E_S^+ \vert ^2E_p^-\\
  \frac{\partial E_S^+}{\partial z } = -\frac{\alpha}{2}E_S^+ + c_B \vert E_p^- \vert ^2E_S^+\\
    \frac{\partial E_S^-}{\partial z } = +\frac{\alpha}{2}E_S^- - c_B \vert E_p^+ \vert ^2E_S^-
    \label{eqn:fieldeq4}
\end{eqnarray}


\noindent where $E_p^{+(-)}$ and $E_S^{+(-)}$ are the forward (backward) propagating pump and Stokes fields, $\alpha$ is the fiber attenuation coefficient and $c_B=\frac{n g_B}{\eta_0}$ is a coefficient proportional to $g_B$, the Brillouin gain, with $n$ being the index of refraction and $\eta_0$ being the impedance of free space. In general, $g_B$ is a function of frequency, however the Brillouin gain spectrum is typically a peak function displaced from the pump frequency by $10$\,GHz, while having a linewidth of $10$\,MHz. This peak is very sharp and we can approximate the Brillouin gain spectrum as a delta function and consider $c_B$ a pump frequency independent constant. Moreover, the frequency of Stokes light and its detuning from cavity is not significant since the cavity is not pumped externally with Stokes light, therefore no interference takes place after one round trip. From the point of view of Stokes light, the cavity is simply two mirrors where a certain fraction of light is lost. The simulation method and parameters are described in Section~\ref{sec: Simulation vs experiment} and the details of the discretisation scheme are found in Appendix~\ref{Appendix: Discretisation}.

Eqs. \ref{eqn:fieldeq1}-\ref{eqn:fieldeq4} only describe SBS and have no source term for the Stokes fields which would correspond to spontaneous Brillouin scattering. It is possible to add a stochastic term to these equations, however this method has several disadvantages. Firstly, including spontaneous scattering would increase the computational complexity of the problem as it would make the equations inhomogeneous. For an accurate calculation, one has to calculate the different mechanical modes of the optical fiber and calculate the phonon occupation numbers in thermal equilibrium~\cite{Laude2018}. One could use a simpler, but still stochastic model known as the distributed fluctuating source model~\cite{PhysRevA.42.5514}, however, its qualitative predictions are similar to the simpler, localised nonfluctuating source model. In this case, SBS is initiated by injecting a very weak Stokes field at the end of the fiber with a random phase. The amplitude of the injected Stokes field is not significant, as long as it is negligible compared to the equilibrium value of the Stokes field. In this case, the injected field is only responsible for initiation of SBS and will not influence the interplay between the pump and Stokes fields significantly. Following the earlier argument on the insignificance of the Stokes frequency and its detuning from cavity resonance, instead of using a random phase, we inject a weak Stokes field that does not interfere with field circulating the cavity, which is the case on average if a field with random phase is injected.

The set of nonlinear equations presented before model SBS only. The interaction between the optical fields and the mechanical oscillator can be described with a simple theoretical model using the typical formalism of optomechanics where we explicitly include the dynamical equation for the mechanical oscillator. Furthermore, we include the thermoptic phase noise in the pump field equation of motion following \cite{Pontin2018}. Based on Eq.~\ref{eqn:fieldeq1}-\ref{eqn:fieldeq4}, the equations of motion for the cavity field $a$ in the frame rotating with the laser frequency, and the Stokes intensity $\vert b  \vert^2=B $ are given by

\begin{eqnarray}
\label{eq:adot}
& \dot{a} =-(\kappa+i(\Delta+\dot{\phi})+iGx)a+\sqrt{2\kappa_{ex}}a_{in}-\frac{G_B}{2}Ba \\
& \dot{B} = -2\kappa B + G_B\vert a \vert^2 B
\label{eq:Bdot}
\end{eqnarray}

\noindent where $\kappa$ is the half-linewidth of the cavity, $\Delta$ is the cavity detuning, $\dot{\phi}$ is the frequency noise with properties outlined in Appendix \ref{Appendix: fiber noise}, $G$ is the cavity resonant frequency shift per displacement, $x$ is the mechanical displacement around the mean position, $\kappa_{ex}$ is the loss rate due to the input mirror and $G_B$ is a constant proportional to the Brillouin gain and can be derived by considering the different normalisations of $E$ and $a$ as shown in Appendix~\ref{Appendix: Normalisations}.  An advantage of this formalism is that the equations of motion can be linearised around a steady state solution and these are analytically solvable. However, this model neglects any spatial variation of the fields since each are described by a single complex amplitude. Moreover, this model considers a single cavity mode only in contrast with the spatial model which considers all modes. In the case of a high Finesse cavity this is a very good approximation but in our case this introduces a small error. There is no stochastic source in Eq.~\ref{eq:Bdot} as we assume that above the Brillouin threshold, the Stokes field always builds up from noise and stimulated scattering will be dominant over spontaneous scattering once equilibrium is reached. We separate both the pump field and the Stokes intensity to a mean value ($\bar{a}$, $\bar{B}$) and to a fluctuating term ($\delta a$, $\delta B$). Neglecting the small mechanical displacement and solving the steady-state equations above Brillouin threshold yields

\begin{eqnarray}
\label{eq:abar}
& \bar{a}=\frac{\sqrt{2\kappa_{ex}}a_{in}}{\kappa-i\Delta+\frac{G_B}{2}\bar{B}} \\
& \bar{B}=\frac{2}{G_B}\bigg(\sqrt{\frac{G_Ba^2_{in}\kappa_{ex}}{\kappa}-\Delta^2}-\kappa \bigg).
\label{eq:Bbar}
\end{eqnarray}

\noindent Since $\bar{B}$ is real and positive, Eq.~\ref{eq:Bbar} is only valid if $\sqrt{\frac{G_Ba^2_{in}\kappa_{ex}}{\kappa}-\Delta^2}>\kappa$ and if $\frac{G_Ba^2_{in}\kappa_{ex}}{\kappa}>\Delta^2$, which indicates the threshold behaviour of Brillouin scattering. An interesting feature of Eqs.~\ref{eq:abar}~and~\ref{eq:Bbar} is that above the Brillouin threshold, the steady-state pump intensity has a constant value of  $\vert a \vert^2 = 2\kappa / G_B$. From the threshold condition, setting the detuning to zero and considering that $G_B \propto \kappa$ (see Appendix \ref{Appendix: Normalisations} it follows that threshold power scales with linewidth as $\kappa^{-2}$. Ignoring fiber losses and keeping the reflectivity of cavity mirrors constant, this means that the threshold power scales as $L^{-2}$ with cavity length as opposed to the $L^{-1}$ scaling for single pass in a fiber. The equations of motion for the fluctuations and for the mechanical displacement are linearised by neglecting any quadratic or higher order terms in fluctuations. We obtain

\begin{eqnarray}
\frac{d(\delta a)}{dt}&= -(\kappa-i(\Delta+\dot{\phi}) - \frac{G_B \bar{B}}{2})\delta a + iG\bar{a}x-\frac{G_B \bar{a}}{2}\delta B\label{eq:dadot}\\
\frac{d(\delta B)}{dt}&=(-2\kappa+G_B \vert \bar{a} \vert^2)\delta B + G_B\bar{B}(\bar{a}^*\delta a + \bar{a} \delta a^*) \label{eq:dBdot}\\
\frac{d^2x}{dt^2}&= -\Omega_m^2 x - \Gamma_m\dot{x}+\frac{\hbar G}{m} (\bar{a}^*\delta a + \bar{a} \delta a^*+\delta B)+\frac{F_{ex}}{m}\label{eq:xddot}
\end{eqnarray}

\noindent where $F_{ex}$ is the sum of all external forces. After Fourier transforming these equations and some algebra, we find an analytic expression for the optomechanical contribution to the inverse mechanical susceptibility, which is given in Appendix~\ref{Appendix: Susceptibility} with derivation.

\section{\label{sec: Simulation vs experiment} Simulation results vs. experiment}

\begin{figure}[h]
\includegraphics[clip, trim={0cm 0cm 0cm 0cm}, width=0.48\textwidth]{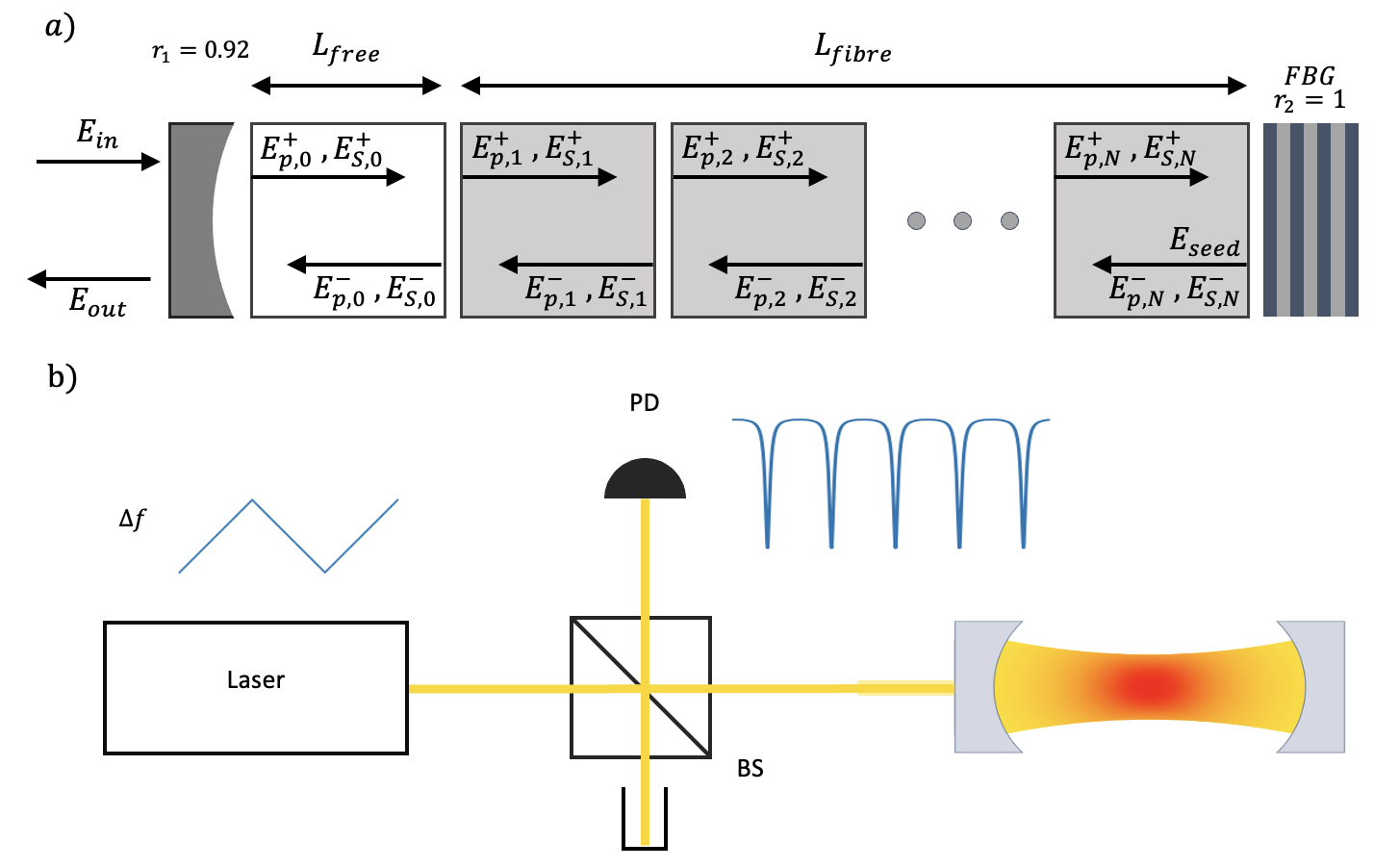}
\caption{a): Schematic of the simulated cavity containing a free space volume of length $\delta z$ and a fiber which is split up in regions of length $\delta z / n$. Forward (backward) propagating pump and Stokes fields are defined at the beginning (end) of each element. b): Schematic of the experiment to measure the finesse of the cavity. The frequency of a $1064$\,nm laser is scanned by applying a ramp signal. The optical power reflected from the cavity is measured with a photodiode and the Finesse is obtained by fitting the measured signal with the Airy-distribution in reflection.}
\label{fig:setup and simulation schematic}
\end{figure}

To verify the model used in the simulations and the theoretical model, we compare the numerical calculations with the cavity finesse measured as a function of input power in our experiment. Here, we focus on the optical aspect of the system only (i.e., no mechanical oscillator). The schematic of the simulated and experimental setup can be seen on Figure~\ref{fig:setup and simulation schematic}. The cavity consists of a standard silica input mirror of reflectivity $0.85$ and a single mode polarisation maintaining optical fiber of length $10$\,m with a fiber Bragg-grating on one end. Between the mirror and the fiber, we use a fixed focus aspheric lens collimator to couple the beam into the fiber. The cavity is driven by a $1064$\,nm laser. A non-polarising beamsplitter is used to collect a fraction of the light reflected from the cavity on a photodetector.

We measure the spectrum of the cavity by modulating the frequency of the laser with a ramp signal scanning $2-3$ free spectral ranges. The simulated finesse is obtained as follows. We consider a fiber cavity with a free space region between the input mirror and fiber input of length of $L_{free}=0.1$ m and a fiber of length $L_{fib} = L_{free} N/n = 10.003$ m, where $N=145$ is the number of finite elements of the fiber using an index of refraction of $n = 1.4496$~\cite{Tan1998}. An input mirror of reflectivity $R_1 = 0.85$ and an output mirror of reflectivity $R_2 = 1$ was used. These parameters are based on the experimental setup. The summary of the parameters and quantities derived from them can be seen in Table\,\ref{Tab: Cavity parameters}.

\begin{table}[]
\begin{tabular}{|c|c|c|}
\hline
\textbf{Parameter} & \textbf{Value} & \textbf{Description} \\ \hline
$L_{free}$                      & 0.1 m            & Free space length of cavity                    \\ \hline
$L_{fiber}$                     & 10.003 m         & fiber length of cavity                         \\ \hline
$N$                             & 145              & Number of finite elements of fiber             \\ \hline
$n$                             & 1.4496           & Refractive index of fiber                      \\ \hline
$R_1$                           & 0.85             & Reflectivity of input mirror                   \\ \hline
$R_2$                           & 1                & Reflectivity of second mirror                  \\ \hline
$\Delta \omega_{FSR}/2 \pi$     & 10.278 MHz       & Free spectral range                            \\ \hline
$\kappa_{L}/2 \pi$              & 436.8 kHz        & Lorentzian half-linewidth of cavity            \\ \hline
$\mathcal{F}_{L}$               & 11.76            & Lorentzian finesse of cavity                   \\ \hline       $\kappa_A/2 \pi$                & 430.8 kHz        & Airy half-linewidth of cavity                  \\ \hline
$\mathcal{F}_A$                 & 11.93            & Airy finesse of cavity                         \\ \hline

\end{tabular}
\caption{Parameters of the cavity used in the simulation. }
\label{Tab: Cavity parameters}
\end{table}

Eqs 1-4 are solved numerically, until the cavity reaches equilibrium at a fixed laser detuning far away from cavity resonance. The frequency of the input field is then swept through one free spectral range at the same rate as in the experiment. The optical power in reflection was recorded during the sweep and the data was filtered in frequency domain with a low pass filter to account for the finite bandwidth of the photodetector used in the experiment. The filtered time-domain data is fitted with the Airy-distribution in reflection~\cite{Bornwolf} allowing to estimate the reflectivity of the back mirror plus the overall optical losses due to absorption in the fiber and mode-matching losses at the fiber-free space interface. The Airy finesse of the cavity is calculated from this effective reflectance and compared to the experimental Airy finesse data. Since the Brillouin gain coefficient and the fiber loss coefficient are not known with a high precision, these are free parameters in matching the simulated data to the experimental ones. Furthermore, a change in the fiber loss coefficient changes the cavity finesse under the Brillouin threshold, which is a known constant. The only free parameter to be varied is the mode matching loss coefficient at the fiber-free space interface, denoted by $\beta$ such that $\beta$ is the ratio of the optical powers after and before the fiber-free space interface. The simulated data is optimised by varying these three parameters using the L-BFGS-B algorithm~\cite{Zhu1997} with the error function being the sum of the squared residuals. The initial guess for the parameters based on coarse manual optimisation and the parameters of the fiber used are $g_B=1.13\times 10^{-11}$\,m$/$W, $\alpha = 5.62 \times 10^{-4}$\,m$^{-1}$ and $\beta = 0.7$. To ensure that the optimisation results are meaningful, we set a factor of 2 relative bound on the Brillouin gain and $\pm 25$ percent relative bound on the fiber and mode-matching losses.

\begin{figure}[h]
\includegraphics[clip, trim={1.25cm 1.75cm 2.8cm 2.5cm}, width=0.48\textwidth]{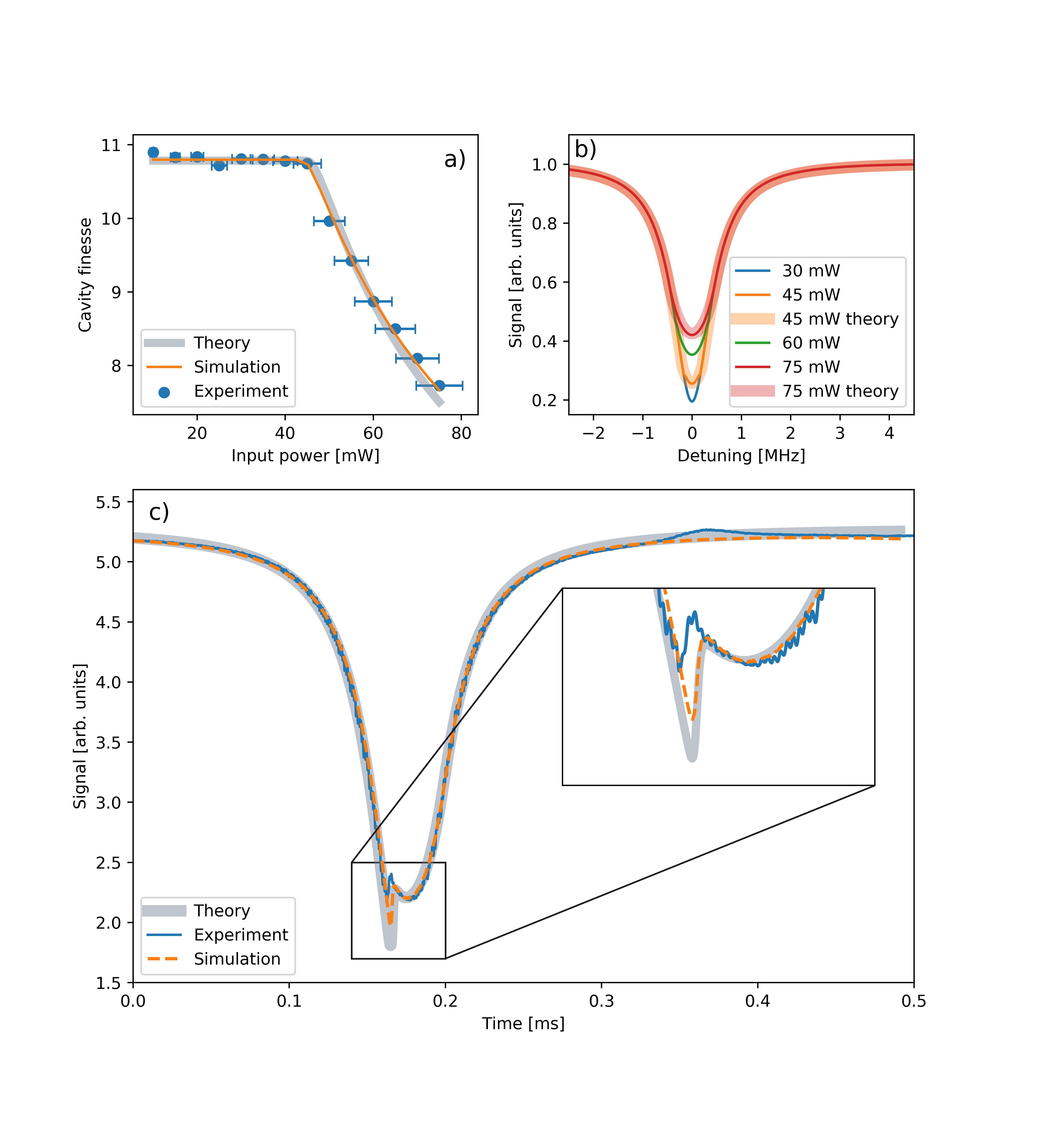}
\caption{\label{fig:fig2} a): Experimental finesse vs input power, with simulation results optimised over the Brillouin gain, fiber loss coefficient and mode-matching loss coefficient. b): Simulated equilibrium cavity spectrum in reflection under Brillouin threshold ($30$\,mW input power) and above Brillouin threshold for $45$\,mW, $60$\,mW and $75$\,mW input powers. c): Dynamic crossing of Brillouin threshold observed as the non-equilibrium cavity spectrum in reflection. The small discrepancy in the experimental curve around $0.4$\,ms is the effect of the small birefringence of the cavity.   }
\end{figure}

The result of the optimisation can be seen on Figure \ref{fig:fig2} a) showing a good agreement between simulation, experiment and theory. To make sure that the finesse under the Brillouin threshold for the theory matches the experimentally measured value, for this specific result, we extended the theory presented in the previous section to include two more cavity modes. Furthermore, we used the three mode theory under the Brillouin threshold to get accurate values for $\kappa$ and $\kappa_{ex}$, found in Table \ref{Tab: Simulation and theory parameters}. The values of these parameters are only well defined in terms of other known constants in the high finesse approximation and there is only a good match between theory and simulation above Brillouin threshold if the accurate values are used. The Brillouin gain coefficient obtained after optimisation is $1.67 \times 10^{-11}$\,m/W. Nikles\,\textit{et al.}~\cite{Nikles1997} measured the Brillouin gain spectrum of optical fibers with different GeO${_2}$ doping levels and obtained peak Brillouin gains varying between $1.63 \times 10^{-11}$\,m$/$W and $5 \times 10^{-11}$\,m$/$W. After correcting for the different wavelengths of light used ($g_B \propto \lambda_p^{-4}$, where $\lambda_p$ is the wavelength of the pumping light) we obtain values between $6.8 \times 10^{-12}$\,m$/$W and $2.1 \times 10^{-11}$\,m$/$W, demonstrating a good agreement between simulation and experiment. Furthermore, the optimal fiber loss coefficient was 12\% higher than the expected value measured by the manufacturer at $1060$\,nm and the mode matching loss coefficient was 4.3\% lower than the initial guess. A summary of the optimisation results can be found in Table~\ref{Tab: Optimised parameters}. Therefore, we can conclude that the simulation was able to reproduce the experimental results with the parameters in agreement with literature or with reasonable differences from the initial estimated values, which can be attributed to measurement uncertainties and measuring fiber parameters at a different wavelength.

For the simulations in the next sections, we used the parameters listed in Table~\ref{Tab: Simulation and theory parameters}. These are a result of an initial manual optimisation for the simulated finesse vs input power results to match the experimental data. Since these parameters are close to the optimisation result, the simulation parameters give qualitatively similar results. The purpose of this paper is to explore the optomechanical effects in a fiber cavity and not to perform precise measurements on the fiber parameters.

\begin{table}[]
\begin{tabular}{|c|c|c|}
\hline
\textbf{Parameter} & \textbf{Value} & \textbf{Description} \\ \hline
$g_B$                      & $1.13 \times 10^{-11}$ m/W           & Brillouin gain                    \\ \hline
$\alpha$                   & $5.62 \times 10^{-4}$ $m^{-1}$       & fiber loss coefficient            \\ \hline
$\beta$                    & 0.70                                 & Mode-matching loss                \\ \hline
$\kappa/2\pi$              & 432.9 kHz                            & Half-linewidth                    \\ \hline
$\kappa_{ex}/2\pi$         & 111.8 kHz                            & Input mirror decay rate           \\ \hline

\end{tabular}
\caption{Parameters used in the simulation and theoretical calculations.}
\label{Tab: Simulation and theory parameters}
\end{table}

\begin{table}[]
\begin{tabular}{|c|c|c|}
\hline
\textbf{Parameter} & \textbf{Value} & \textbf{Description} \\ \hline
$g_B$                      & $1.67 \times 10^{-11}$ m/W            & Brillouin gain                    \\ \hline
$\alpha$                   & $6.31 \times 10^{-4}$ $m^{-1}$                     & fiber loss coefficient            \\ \hline
$\beta$                    & 0.67                                 & Mode-matching loss                \\ \hline

\end{tabular}
\caption{Results of the optimisation for the fiber parameters}
\label{Tab: Optimised parameters}
\end{table}

The experiment demonstrated a dynamic phenomenon when the Brillouin threshold is crossed. When the intracavity power is just above the Brillouin threshold, it takes a long time for the cavity to reach equilibrium as the exponent in the solution of Eqs.~\ref{eqn:fieldeq1}-\ref{eqn:fieldeq4} (ignoring pump depletion effects) is small. This effect is shown on Figure~\ref{fig:fig2}~c) as ripple on the cavity response when the cavity is in a transient regime between a state without SBS to a state with SBS. The same phenomenon is present in the simulation results and agrees well qualitatively with the experimental data.

A second simulation used a different approach to obtain the spectrum of the cavity. Instead of sweeping the frequency of the laser, which is experimentally the easiest way to perform the measurement, the reflected intensity was recorded in equilibrium for different detunings, always starting from an empty cavity. This allows to observe the true equilibrium response of the fiber cavity. The results on Figure~\ref{fig:fig2}~b) show that the equilibrium response of the fiber cavity changes above the Brillouin threshold. If the detuning is large enough such that the intracavity power is under the Brillouin threshold, there is no difference between a normal and a fiber cavity and the cavity spectrum is identical to the Airy-distribution as expected. However, once the detuning is small enough to induce SBS, the intracavity pump power is reduced, while the Stokes power is increased. Since SBS is initiated by spontaneous scattering with a random phase, the Stokes and pump fields simply add up on average and the optical power reflected from the cavity is increased. The resonance lines obtained are therefore effectively broader than the original lines (as well as being different from the Airy-distribution). Therefore, the method used above to extract the finesse in the simulation and experiment is only valid and meaningful for weak Brillouin scattering, where the line shapes are not altered significantly. \\

\section{\label{sec: enhanced cooling} Brillouin scattering enhanced cooling}

\noindent Brillouin processes in optical fibers have being studied for a long time, and have nowadays many technological applications, from all-fiber Brillouin lasers~\cite{brillouin_laser,brillouin_laser_2}  to fiber Brillouin amplifiers~\cite{brillouin_ampli_1,brillouin_ampli_2}. However, in the context of cavity optomechanics, Brillouin processes have only been explored in the direct case, i.e., when the mechanical mode emerges directly from confined Brillouin scattering. This is the typical case for whispering gallery optomechanical systems~\cite{brill_optomech_2_vanner} and for nanophotonic structures in general~\cite{brill_optomech_1}. Here, we propose and demonstrate that under certain conditions SBS in a fiber cavity adds an extra modulation to the intracavity pump intensity, reducing the optical spring effect and increasing the optomechanical cooling rate. Contrary to what previously explored, acoustic phonons constitute a silent partner in the optomechanical interaction.

To simulate this effect with the coupled amplitude equations (Eqs.~\ref{eqn:fieldeq1}-\ref{eqn:fieldeq4}), we consider a fiber cavity with parameters found in Table~\ref{Tab: Cavity parameters}. We also compare the results with the linearised theory with parameters found in Table~\ref{Tab: Simulation and theory parameters}. For the initial search for the optimal parameters, the mechanical motion of the mirror is not simulated, it is explicitly given as a function of time as a pure sinusoidal oscillation with a thermal amplitude $\Delta x_{RMS} = (k_BT/m\Omega_m^2)^{\frac{1}{2}}$ with the temperature of the bath being $T=300$\,K and the mass of the harmonic oscillator being $m=10^{-10}$\,kg, which is typical for high quality factor silicon-nitride membranes~\cite{Chakram2014}. This approximation is valid if the mechanical quality factor of the oscillator is high, which is typical and favourable in experiments to achieve low effective temperatures. After the cavity has been brought to equilibrium, the radiation pressure force on the input mirror is recorded and demodulated in phase and in quadrature with mechanical motion yielding the optomechanical cooling rate and mechanical resonance frequency shift. A fine sweep over detuning is performed (typically $400$ points in a range of $2$\,MHz) as well as a coarse sweep over input power (from $10$\,mW to $300$\,mW in steps of $10$\,mW). For comparison, the simulation is also executed without Brillouin scattering by setting the Brillouin gain to zero. Throughout this section, we consider the main results to be the outcome of the numerical simulation of the coupled amplitude equations with explicit mechanical motion, and compare them to the predictions of the linearised theory. The coupled amplitude equations include nonlinear phenomena as well as spatial dependence of the fields and comparison with theory serves as a benchmark of our theoretical model. Moreover, we compare a single simulation result with explicit mechanical motion to a more detailed, computationally expensive simulation, which treats the stochastic mechanical oscillator dynamically.

The results of the simulation showed that the system exhibits three different qualitative behaviours depending on the input parameters. The first, trivial case, occurs when the intracavity power at zero detuning is under the Brillouin threshold, in this case there is no difference between a cavity with and without SBS. If the peak intracavity power is above the Brillouin threshold, and the frequency of the mechanical oscillator is high compared to the time scales of fluctuations of the Stokes intensity, one can see a destructive effect due to Brillouin scattering both for the optical spring effect and optomechanical cooling as seen on Figure \ref{fig:fig2} a) where we considered a mechanical resonance frequency of $\Omega_m /2\pi = 300$\,kHz. For such mechanical resonance frequencies, the dynamics of Brillouin scattering is too slow, hence the force oscillating at the mechanical resonance frequency is reduced. The results from the linearised theory are in good agreement with the simulation for the expected parameters.

\begin{figure}[h]
\includegraphics[clip, trim={0cm 0.0cm 0cm 0cm}, width=0.48\textwidth]{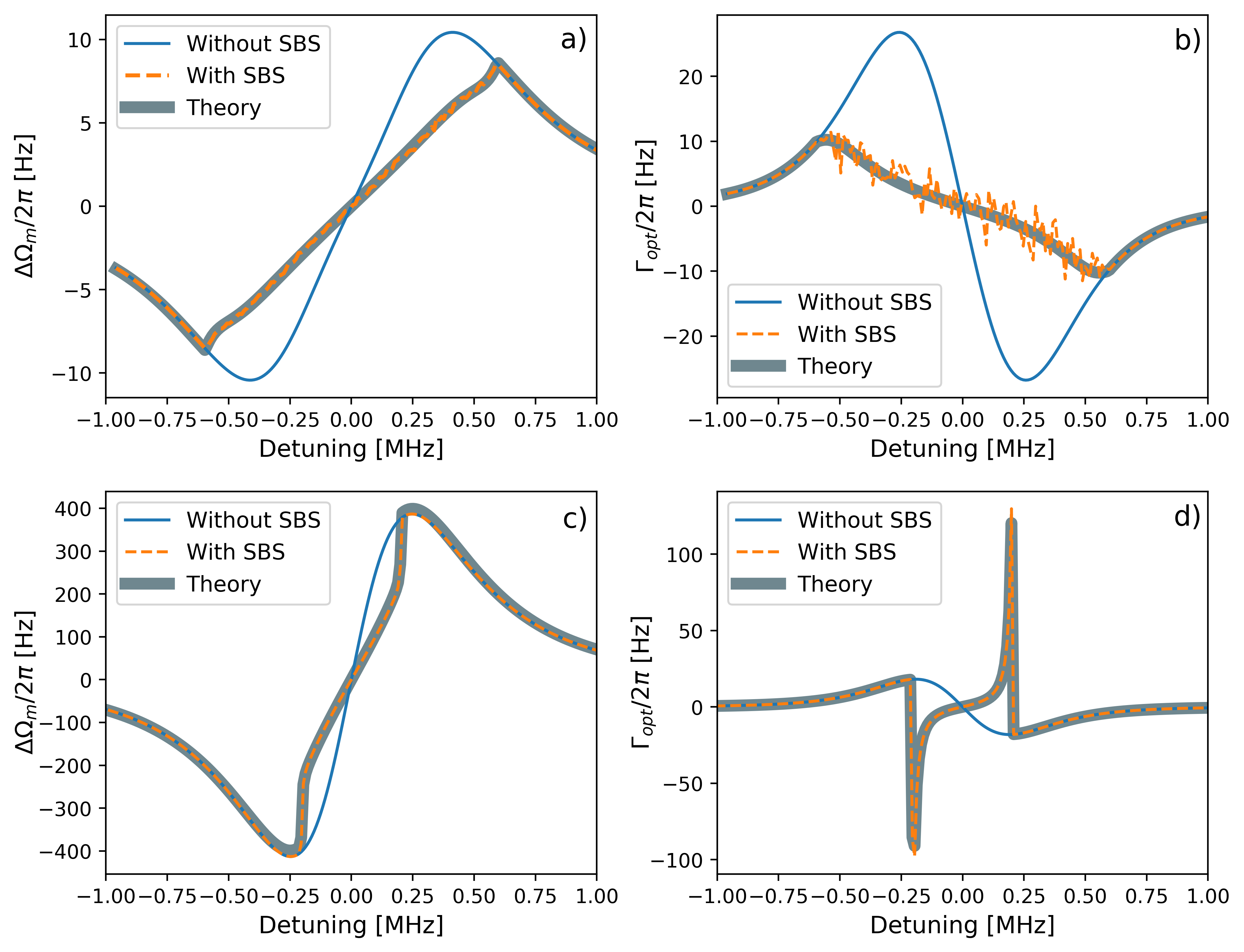}%
\caption{\label{fig:fig3} Optomechanical effects with and without SBS including predictions of linearised theory in a): second regime, mechanical resonance frequency shift,  b): second regime, optomechanical damping, c): third regime, mechanical resonance frequency shift, d): third regime, optomechanical damping.}
\end{figure}

For low mechanical resonance frequencies (up to $\approx 30$\,kHz with the parameters in consideration), the mechanical motion is slow enough such that for a certain detunings, the intensity of the Stokes field oscillating at the mechanical resonance frequency is significant. As a result, the pump field will be modulated stronger than usual, which results in a larger optical damping rate. This effect does not increase the magnitude of the force acting on the mechanical oscillator because Brillouin scattering conserves energy, this was verified by calculating the total force and comparing it to the no SBS case. However, it is capable of introducing a phase shift, decreasing the optical spring effect and increasing optomechanical cooling. This can be seen on Figure~\ref{fig:fig2} b) where the mechanical resonance frequency is $\Omega_m /2\pi = 6.1$\,kHz. As the Brillouin threshold is reached, the optomechanical cooling rate $\Gamma_{opt}$ increases sharply for blue detuning, while the optical spring effect reduces.
The results from the linearized theory are in very good agreement with the simulation results within the accuracy of the approximations used.


\begin{figure}[h]
\includegraphics[clip, trim={0cm 0.0cm 0cm 0cm}, width=0.48\textwidth]{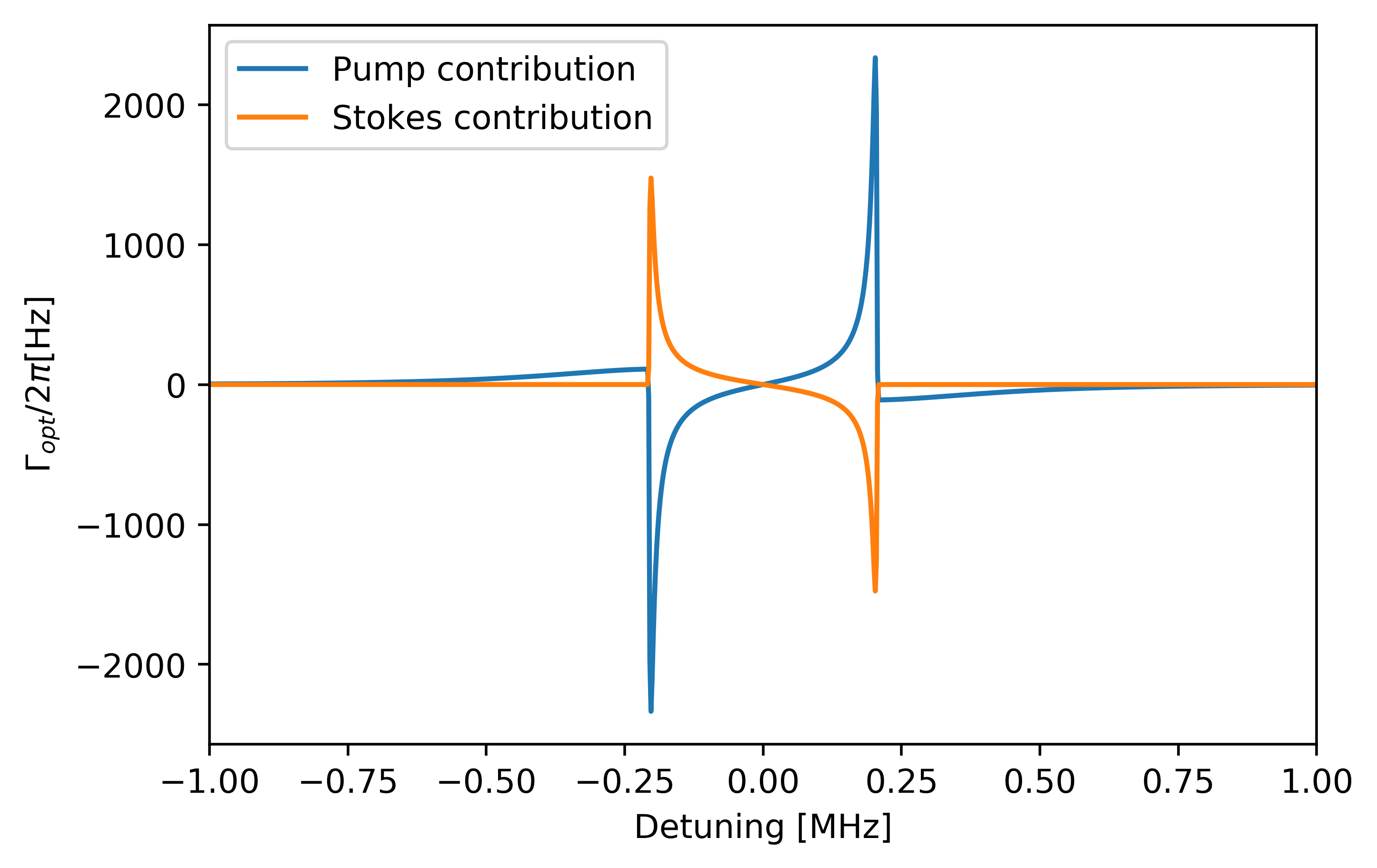}
\caption{\label{fig:two_sources} Optomechanical damping rate due to pump and Stokes fields can be separated since the oscillator will not respond to the $\approx$GHz beatnote. The sharp increase/decrease in the total damping rate is due to the modulation of the pump field by the Stokes field. }
\end{figure}

The origin of the sharp heating peak on the red detuned side can be explained intuitively as follows. In a normal optomechanical system where the drive is red-detuned, as the mechanical oscillator moves in the positive direction it reduces the detuning. This increases the intracavity power which pushes the oscillator back. If we include Brillouin scattering, as the mechanical motion increases the intracavity power, the Brillouin threshold is crossed, the pump intensity is reduced. This allows the mechanical oscillator to reach a larger displacement. The reduced intensity is converted to Stokes intensity, which acts later on the oscillator due to the finite time needed for the system to reach equilibrium. A similar explanation holds for the blue detuned side where the effect is opposite, giving rise to cooling. Figure~\ref{fig:two_sources} justifies this explanation, showing that the extra cooling or heating is due to the additional modulation in the pump field.

\begin{figure}[h]
\includegraphics[clip, trim={0cm, 0cm, 0cm, 0cm},width=0.48\textwidth]{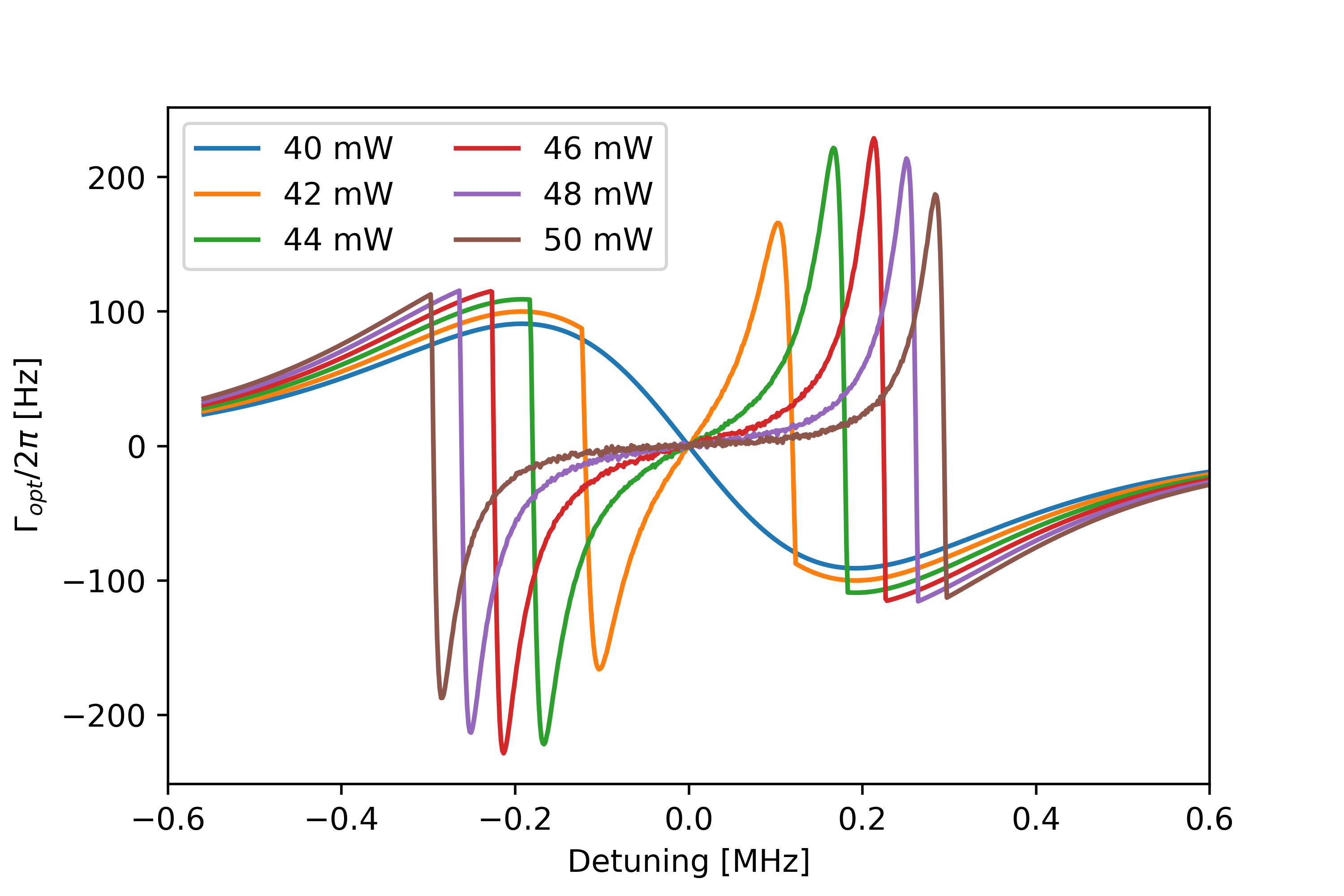}
\caption{\label{fig:Powerplot} Optomechanical damping rate for various input powers. As the input power is increased, the Brillouin threshold is crossed at larger detunings. The maximum optical damping occurs at approximately the input power where the Brillouin threshold is crossed at the minimum of the regular optomechanical damping rate, because the mechanical motion modulates the intracavity power the strongest in this case.}
\end{figure}

To investigate how the Brillouin scattering enhanced cooling depends on various parameters, we change the cavity input power, the mass of the oscillator and the mechanical resonance frequency. The results for different input powers can be seen in Figure~\ref{fig:Powerplot}. For this comparison, we consider the typical system with $\Omega_m/2\pi = 20$\,kHz. By increasing the input power, the detuning corresponding to the Brillouin threshold changes, as well as the time for the Stokes field to reach equilibrium, hence the phase shift between mechanical motion and Stokes intensity. Moreover, the modulation amplitude of the Stokes field depends on the modulation amplitude of the pump field, which is maximum approximately at a detuning of $\frac{\sqrt{3}\kappa}{6 }= 2\pi \times 289 $ kHz considering a Lorentzian response. These two effects result in an optimal cavity input power of approximately $46$\,mW  which maximises the optomechanical damping rate.

\begin{figure}[h]
\includegraphics[clip, trim={0cm, 0cm, 0cm, 0cm},width=0.48\textwidth]{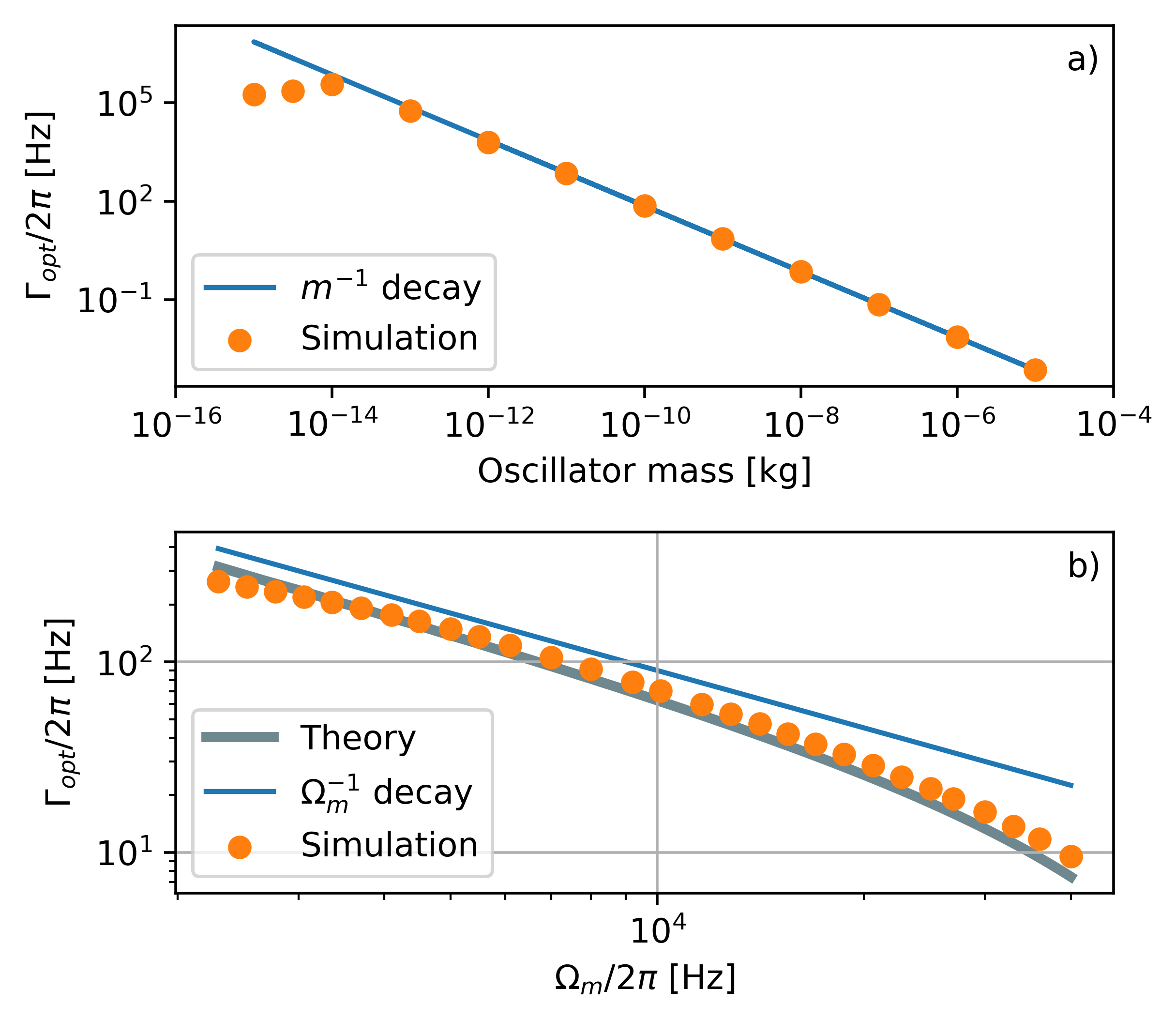}
\caption{\label{fig:fig4} Maximum optomechanical damping rate due to Brillouin scattering as a function of mass of harmonic oscillator (a) and as a function of mechanical resonance frequency (b). The $\Omega_m^{-1}$ decay plot is a guide to the eye only. }
\end{figure}

To see the effects of the mass of the oscillator, we considered the system described in Sec.~\ref{sec:Theory}, with $\Omega_m/2\pi = 10$\,kHz and input power of $60$\,mW and we find the maximum optomechanical damping rate as a function of detuning. The peak optomechanical damping rate due to Brillouin scattering decays as $m^{-1}$ for a broad range of masses just as the linearised theory with and without Brillouin scattering predicts. For very low masses, the mechanical amplitude is too large for the linear approximation to be valid.
Similarly, when considering the effect of mechanical resonance frequency, we considered the system with usual parameters and with input power of $60$\,mW, and extracted the peak optomechanical damping rate due to Brillouin scattering. The results (Figure\,\ref{fig:fig4} b).) show that the damping rate is increasing with decreasing frequency, with a decreasing rate. The previous result for different masses allows us to correct for the different mechanical amplitudes at different mechanical resonance frequencies, giving an optimum around $6$\,kHz, where the gradient of the simulated data matches the gradient of $\Omega_m^{-1}$ decay predicted by the linearised theory without Brillouin scattering. The results from the linearised theory are in good agreement with the simulation.

\begin{figure}[h]
\includegraphics[clip, trim={0.5cm, 1.10cm, 1.55cm, 2cm},width=0.48\textwidth]{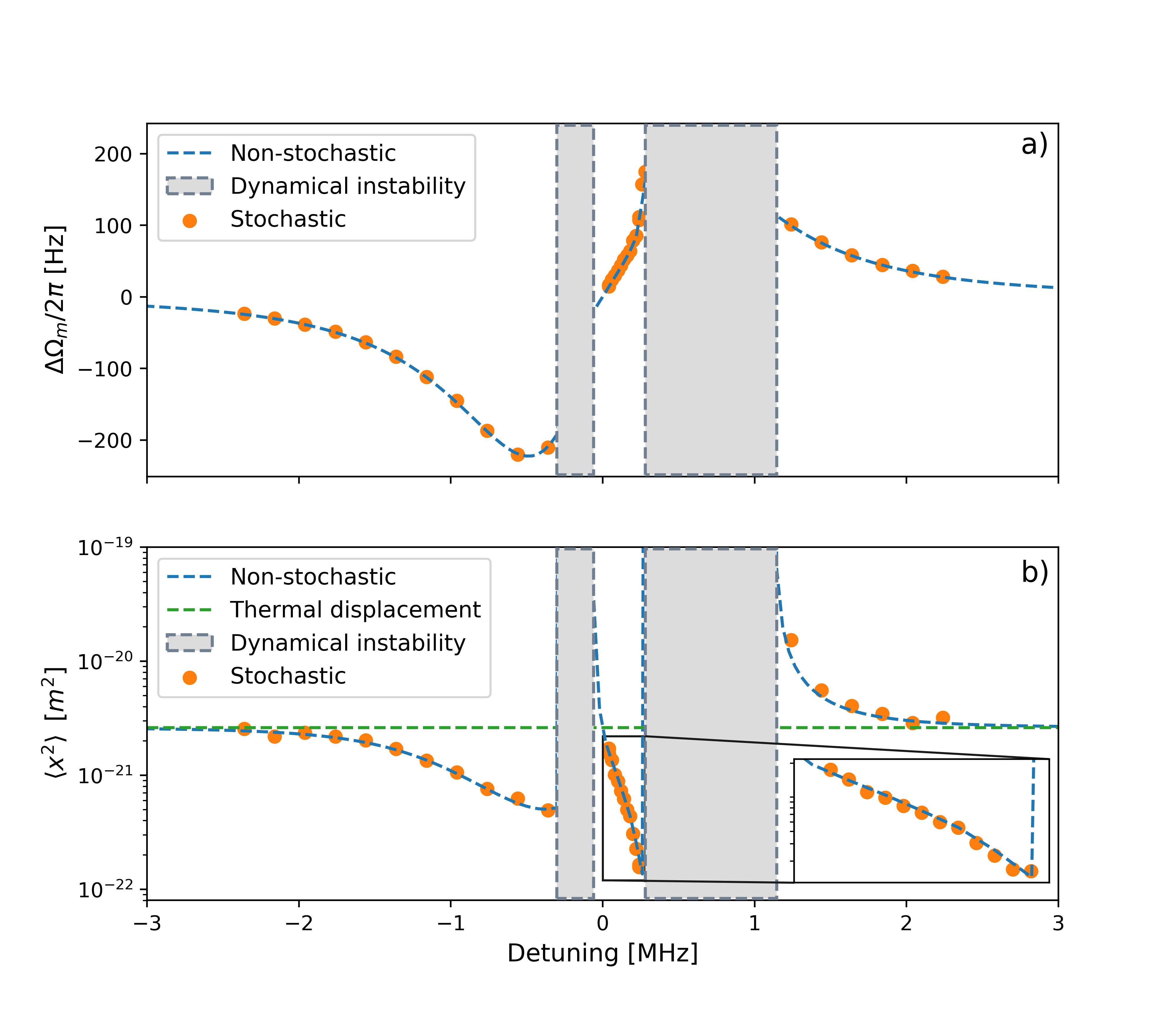}
\caption{\label{fig:fig5} a): Mechanical resonance frequency shift from non-stochastic and stochastic simulations. b): Mean squared mechanical displacement predicted from non-stochastic simulation vs. result from stochastic simulation. Inset: high resolution results where maximum cooling is expected. In both figures, the region where the negative optomechanical damping coefficient is larger than the mechanical linewidth is shaded.}
\end{figure}

Even though the previous simulation method was able to find the region in the parameter space where Brillouin scattering enhances optomechanical cooling, it neglected the two-way coupling between the cavity and the mechanical motion which affected the cavity field but radiation pressure had no effect on the oscillator. Therefore, we numerically solved the fully coupled dynamical equations with no change on the cavity side but with the harmonic oscillator also driven by a Langevin thermal force at a bath temperature of $300$\,K.  The equation of motion was solved with the modified leapfrog method suggested by Mannella~\cite{Mannella2004} ensuring that the uncertainty in position and momentum does not grow over time. Since the time step of the simulation is set by the discretisation of space which was chosen to be $0.1$\,m, the time step was $0.33$\,ns, which in general yields in a very large number of simulation steps required, making the simulation computationally expensive. As a consequence, the length of the fiber was reduced to $L_{fib} = 5$\,m, the frequency of the mechanical oscillator was $\Omega_m / 2\pi = 20$\,kHz and the quality factor was $Q=5000$ giving $\Gamma_m/2\pi = 4$\,Hz. The simulation was ran for $t=0.6$\,s which is approximately $15\Gamma_m^{-1}$. Since the simulation involved a stochastic force, the results were averaged over $50$ different realisations of the noise. The results can be seen on Figure~\ref{fig:fig5}. One can see that the results for both the mechanical resonance frequency shift and the optomechanical cooling are in good quantitative agreement with the previous, simplified simulation. Only in the regime of mechanical lasing, where the optomechanical damping rate is negative and larger than the mechanical linewidth, the results from the two simulations deviate, but this is expected as the simplified simulation did not involve the dynamics of the mechanical oscillator.

\section{\label{sec: Ground state cooling} Ground state cooling with a fiber cavity}

To investigate whether ground state cooling is possible in a fiber cavity exploiting Brillouin scattering, we use the equations of motion approach to find the mean number of phonons (see Appendix~\ref{Appendix: Phonon number} for details) and the linearised theory. To suppress thermal and thermoptic noise, we considered the mechanical oscillator and the optical fiber in a liquid nitrogen cryogenic environment at 77\,K. At cryogenic temperatures, the parameters of the fiber noise spectral density (see Appendix \ref{Appendix: fiber noise} and the peak Brillouin gain are different from the room temperature value. Instead of peak Brillouin gain, previous experiments measured the Brillouin linewidth. However, it was shown that the product of peak Brillouin gain and Brillouin linewidth remains constant~\cite{Garcus1997, LEFLOCH2003395}, therefore we can rescale the previously quoted gain by the ratio of linewidths at room temperature and at cryogenic temperatures. Thus we consider $g_B=6.71 \times 10^{-12} \hspace{3pt} m/W$ at 77\,K.

To obtain the minimum phonon number we considered fiber lengths of 0.1 m, 0.2 m, 0.4 m, 0.8 m and 1.6 m. Above 1.6 m, the free spectral range of the cavity becomes comparable with the Brillouin linewidth and the single Stokes mode theory becomes inaccurate. We minimised computationally the mean number of phonons by varying the ratio of input power and Brillouin threshold power, cavity finesse and mechanical resonance frequency oscillator. The finesse is bounded between 20 and 150 to ensure that the single-mode model is appropriate and that the maximum theoretical finesse with fiber cavities is not surpassed~\cite{Pontin2018}. The mass of the mechanical oscillator is chosen to be $10^{-12}$\,kg and the quality factor is $10^9$. We calculated the optomechanical damping rate as a function of detuning using the linearised theory in the $[0, 2\kappa]$ interval. Next, we estimated the final number of thermal phonons for the largest optomechanical damping rate. In case the maximum optomechanical damping rate is negative or the thermal phonon number estimate is more than 100, the total phonon number is set to a large number to avoid unstable and not interesting regions in the parameter space. Otherwise, the phonon number is calculated for 100 detunings $\pm 0.1 \kappa$ around the maximum optomechanical damping rate and the minimum phonon number is picked. Since the obtained number of phonons are often inadmissible due to negative optomechanical damping rates, the error function is not well-behaved and gradient-based optimisation algorithms failed. Therefore, we used the differential evolution algorithm to find the minimum~\cite{Storn1997}. The lowest phonon number of 0.44 is achieved for a fiber length of 1.6\,m, finesse of 150 and mechanical resonance frequency of 12.6\,kHz. Such a low frequency, high quality factor oscillator is not common among clamped oscillators, this could be realised in a levitated optomechanical system. However, the coupling between the cavity field and the oscillator is different in this case and our model is not applicable in this scenario. For more realistic, higher frequency oscillators, it is still possible to reach the ground state using shorter fiber cavities. For a fiber cavity length of 0.1 and mechanical resonance frequency of 105\,kHz, a phonon number of 0.83 was reached.

\section{Conclusion\label{sec: Conclusion}}

In this paper, we have investigated the role of stimulated Brillouin scattering in a fiber cavity optomechanical system. We introduced a set of coupled amplitude equations describing the optical fields in a fiber cavity, and presented a simple theoretical model using the formalism of quantum optics. The latter can be solved analytically after linearisation around a steady-state solution. We compared our experimental results on the optical subsystem with the numerical solutions of the coupled amplitude equations and the steady-state solution of the theoretical model and found a very good qualitative agreement between the three results. We used the coupled amplitude equations with explicit mechanical motion to investigate the possibility of increased optomechanical damping rate due to stimulated Brillouin scattering in the fiber cavity. We found that, for mechanical resonance frequencies up to 30\,kHz, optomechanical cooling was enhanced for blue-detuned driving of the cavity near the Brillouin threshold. The results are in agreement with the linearised theory and a numerical simulation where the motion of the mechanical oscillator was treated dynamically, including a stochastic driving force. Finally, using the linearised theory, we explored the possibility of cooling a mechanical oscillator to average phonon numbers below unity. The results show that it is possible to cool a mechanical oscillator near the ground state for short fiber lengths. The theoretical model and simulations presented in this paper can also be used to describe a cavity which is driven by a second laser which is detuned by the Brillouin shift with respect to the cooling laser. Being able to change the detuning between the Stokes light and the cavity resonance might result in enhanced performance and cooling below mean phonon numbers of unity.

\appendix

\section{Discretisation scheme}
\label{Appendix: Discretisation}
As an initial approximation, we assume that the field amplitudes do not change significantly when propagated over $\delta z$, the distance between two points in discretised space. In this case, we can approximate the solution for \ref{eqn:fieldeq1} (and the other equations) within $\delta z$ as
\begin{equation}
    E_p^{+(0)}(z+\delta z) = E_p^+(z)\Big(1-\frac{\alpha \delta z}{2}-c_B \vert E_S^-(z+\delta z) \vert^2 \delta z\Big)
\end{equation}{}

\noindent where we decided to evaluate $E_S^-$ at $z+\delta z$ due to the backwards propagation direction and the (0) superscript indicates the approximation. Similarly, we can approximate the backwards propagating Stokes field at $z$ as

\begin{equation}
    E_S^{-(0)}(z) = E_S^+(z+\delta z)\Big(1-\frac{\alpha \delta z}{2}+c_B \vert E_p^-(z) \vert^2 \delta z\Big).
\end{equation}{}

To improve this first order approximation, we use the above approximations to account for the spatial change of the fields as they propagate the distance $\delta z$ by defining $E_p^{+}(z+\frac{1}{2}\delta z) = \frac{1}{2}(E_p^+(z) + E_p^{+(0)}(z+\delta z))$ and $E_S^{-}(z+\frac{1}{2}\delta z) = \frac{1}{2}(E_S^-(z+\delta z) + E_S^{-(0)}(z))$. Finally, we use these fields defined halfway between two grid points to get the final equations.

\begin{equation}
    E_p^{+}(z+\delta z) = E_p^+(z)\Big(1-\frac{\alpha \delta z}{2}-c_B \vert E_S^-(z+\frac{1}{2}\delta z) \vert^2 \delta z\Big)
\end{equation}{}
\begin{equation}
    E_S^{-}(z) = E_S^-(z+\delta z)\Big(1-\frac{\alpha \delta z}{2}+c_B \vert E_p^+(z + \frac{1}{2}\delta z) \vert^2 \delta z\Big).
\end{equation}{}

The equations for $E_p^-$ and $E_S^+$ are obtained as a straightforward extension of these equations. After each cavity round-trip and when reflected on the moving mirror, the field amplitudes are multiplied by the appropriate complex phase factor that depends on the mechanical displacement. Moreover, we include an additional loss at the free space-fiber interface that accounts for the mode-matching losses which was chosen to be a loss of 30\,\% in optical power, and rescale the amplitudes by the ratio of the beam waist diameter in free space (1\,mm) and the mode-field diameter in the fiber (6.6\,$\mu$m).

\section{\label{sec}Properties of thermoptic noise}
\label{Appendix: fiber noise}
The fiber thermoptic noise is characterised by the power spectral density

\begin{equation}
    S_{\dot{\phi}\dot{\phi}}(\omega) = \frac{\pi c^2 k_B T^2 q^2}{4 \kappa_t \lambda^2 L} F(\omega)
\end{equation}

\noindent where

\begin{equation}
    F(\omega) = \ln{\Big( \frac{k_{min}^4+(\omega/D)^2}{k_{max}^4 + (\omega/D)^2} \Big)}.
\end{equation}

In this expression, $q=\alpha+\frac{1}{n}\frac{\partial n }{\partial T}$ where $\alpha$ is the linear expansion coefficient, $k_t$ is the thermal conductivity, $k_{max} = 2/w_0$, $k_{min} = 2/a_f$ where $w_0$ and $a_f$ are the fiber mode-field and fiber outer radii respectively, and $D$ is the thermal diffusivity. As some of these parameters depend on temperature, their values are summarised in Table \ref{Tab: fiber noise parameters}. The $q$ coefficient can be experimentally measured with fiber Bragg gratings and we used the results of Reid and Ozcan for the non-embedded United Technologies fiber \cite{Reid1998}. The thermal conductivity was estimated by interpolating the data from Brown \cite{Brown2005}. We assume that the fiber mode field and fiber outer radii do not change significantly the noise spectral density due to the low thermal expansion coefficient and the logarithmic dependence. The thermal diffusivity was calculated as $D=k_t/(c_p \rho)$ where $c_p$ is the specific heat capacity at constant pressure and $\rho$ is the mass density. Due to the low thermal expansion coefficient of fused silica, we only considered the temperature dependence of $k_t$ and $c_p$. At room temperature, we used a measured value \cite{Weber2003}. At cryogenic temperatures, $c_p \approx c_v$ and used the Debye model with a Debye temperature of 500 K \cite{Nakayama2002}. The results are shown in Table \ref{Tab: fiber noise parameters}.
\begin{table}[h]
\begin{tabular}{|c|c|c|}
\hline
\textbf{Parameter}                         & \textbf{300 K}         & \textbf{77 K}          \\ \hline
$q \hspace{3pt} [K^{-1}]$                  & $6.7\times 10^{-6}$    & $2.8\times 10^{-6}$    \\ \hline
$\kappa_t \hspace{3pt} [Wm^{-1}K^{-1}]$    & $1.35$                 & 0.52                   \\ \hline
$D \hspace{3pt} [m^2s^-1]$                 & $8.2\times 10^{-7}$    & $2.53\times 10^{-6}$   \\ \hline
\end{tabular}
\caption{Parameters used for phonon number calculations at room and cryogenic temperatures. }
\label{Tab: fiber noise parameters}
\end{table}
\section{\label{sec}Normalisations}
\label{Appendix: Normalisations}

To find the constant of proportionality between the spatial Brillouin gain $c_B$ used in the simulation and the temporal Brillouin gain $G_B$ used in theory, we have to consider the different normalisations used. The field amplitudes are normalised such that the time-averaged energy density corresponding to each forward/backward propagating travelling wave in the fiber is

\begin{equation}
    \langle u \rangle = \frac{n^2\epsilon_0}{2} \vert E^2 \vert.
\end{equation}

The total energy of electromagnetic radiation in the cavity is then

\begin{equation}
    E_{cav} = \langle u \rangle V_{fib}
\end{equation}

\noindent where $V_{fib}$ is the fiber volume where light propagates. Since we consider a one dimensional model only, we take the light intensity to be constant at a given cross section of the fiber, $V_{fib} = L_{fib}\pi(3.3\times 10^{-6} m)^2$ since the mode-field diameter of the fiber is 6.6 $\mu m$. The number of photons in the cavity is then $n_{cav} = E_{cav}/\hbar \omega_L$, which is by definition equal to $\vert \langle a \rangle \vert ^2 $. Using the fact that in the fiber, $\partial / \partial z = (n/c) \partial / \partial t$ and comparing Eqs. \ref{eqn:fieldeq1} and \ref{eq:adot}, we arrive at the relation

\begin{equation}
    G_B = \frac{2 \hbar \omega_L c}{V_{fib} \epsilon_0 n^3}c_B.
\end{equation}

\section{\label{sec}Optomechanical susceptibility}
\label{Appendix: Susceptibility}

After Fourier transforming Eq. (\ref{eq:dBdot}) and solving for $\delta\hat{B}$, we get

\begin{equation}
    \delta\hat{B}[\omega] = \frac{iG_B\bar{B}}{\omega}(\bar{a}^*\delta \hat{a}[\omega] + \bar{a} \delta \hat{a}^*[\omega])
    \label{eq:dBomega}
\end{equation}

\noindent where hats and square brackets denote Fourier transforms of the dynamical variables. Similarly, from Fourier transforming Eq. (\ref{eq:dadot}) and grouping all terms containing $\delta \hat{a}[\omega]$, we get

\begin{equation}
    f(\omega)\delta \hat{a}[\omega]=iG_B\bar{a}\hat{x}[\omega]-g(\omega)\delta \hat{a}^*[\omega]
\end{equation}

\noindent where

\begin{equation}
    f(\omega)=\kappa+\frac{G_B\bar{B}}{2}-i(\omega+\Delta-\frac{G_B^2\bar{B}\vert\bar{a}\vert^2}{2\omega})
\end{equation}

\noindent and

\begin{equation}
    g(\omega) = -\frac{iG_B^2\bar{B}\bar{a}^2}{\omega}.
\end{equation}

As usual, $\delta \hat{a}^*[\omega]=(\delta\hat{a}[-\omega])^*$, therefore it is possible to find the following expression that relates $\delta \hat{a}[\omega]$ and $\hat{x}[\omega]$:

\begin{equation}
    \delta \hat{a}[\omega]= \frac{iG_B\bar{a}-\frac{iG_B\bar{a}^*g(\omega)}{f^*(-\omega)}}{f(\omega)h(\omega)}\hat{x}[\omega]
    \label{eq:daomega}
\end{equation}{}

\noindent where

\begin{equation}
    h(\omega) = 1-\frac{g(\omega)g^*(-\omega)}{f(\omega)f^*(-\omega)}.
\end{equation}

Combining the Fourier transform of Eq. (\ref{eq:xddot}), Eq. (\ref{eq:dBomega}) and Eq. (\ref{eq:daomega}), then solving for the optomechanical contribution to the inverse susceptibility yields

\begin{equation}
    \chi_{opt}^{-1}(\omega)=-\frac{\hbar G_B}{\hat{x}[\omega]} \Big(1+\frac{iG_B\bar{B}}{\omega}\Big)(\bar{a}^*\delta \hat{a}[\omega]+\bar{a}\delta \hat{a}^*[\omega])
\end{equation}{}

\noindent from which we obtain the optomechanical damping rate and mechanical resonance frequency shift.

\section{\label{sec} Mean number of phonons}
\label{Appendix: Phonon number}
To find the mean number of phonons, we use the equations of motion approach. For this, it is convenient to rewrite the equations of motion for the Stokes amplitude $b$ instead of the Stokes intensity ($B$). The goal is to write the Fourier transform of the position observable as

\begin{equation}
\begin{split}
    \hat{x}[\omega] = \chi_{eff}(\omega)(\hat{F}_{th}[\omega]+\chi_a(\omega)\hat{a}_{in}[\omega]+\chi_b(\omega)\hat{b}_{in}[\omega]+\\
    \chi_a^*(-\omega)\hat{a}^\dagger_{in}[\omega]+\chi^*_b(-\omega)\hat{b}^\dagger_{in}[\omega] + \chi_{\dot{\phi}}(\omega)\dot{\phi}(\omega))
\end{split}
\end{equation}

\noindent where the three quantum noise terms have the following correlation functions

\begin{eqnarray*}
    \langle \hat{F}_{th}(t)\hat{F}_{th}(t')\rangle&=&2k_BTm\Gamma_m\delta(t-t') \\
    \langle \hat{a}_{in}(t)\hat{a}^\dagger_{in}(t')\rangle&=&\langle \hat{b}_{in}(t)\hat{b}^\dagger_{in}(t')\rangle=\delta(t-t') \\
    \langle \hat{a}^\dagger_{in}(t)\hat{a}_{in}(t') \rangle&=&\langle \hat{b}^\dagger_{in}(t)\hat{b}_{in}(t') \rangle =0
\end{eqnarray*}

\noindent and the power spectral density of the detuning noise is given in Appendix \ref{Appendix: fiber noise}.
Using these correlators, the power spectral density of the position observable becomes

\begin{multline*}
    S_{xx}(\omega) = |\chi_{eff}(\omega)|^2(2k_BTm\Gamma_m+|\chi_a(\omega)|^2+|\chi_b(\omega)|^2 \\
    + S_{\dot{\phi}\dot{\phi}}(\omega) |\chi_{\dot{\phi}}(\omega)|^2 ).
\end{multline*}

\noindent The optical transfer functions are

\begin{equation}
\begin{split}
    &\chi_a(\omega) = \hbar G \sqrt{2\kappa}\Big( 1+\frac{iG_B \vert \bar{b} \vert^2}{\omega}\Big)(\frac{\bar{a}^*}{f(\omega)h(\omega)}+ \\
    &\frac{\bar{a} g^*(-\omega)}{h^*(-\omega)f^*(-\omega)f(\omega)})
\end{split}
\end{equation}

\begin{equation}
\begin{split}
    \chi_b(\omega) = \frac{-i\hbar G \sqrt{2\kappa} G_B \bar{b}^*}{2\omega}\Big( 1+\frac{iG_B \vert \bar{b} \vert^2}{\omega}\Big)(\frac{\vert a \vert^2 }{f(\omega)h(\omega)}+ \\
    \frac{\vert a \vert^2 }{f^*(-\omega)h^*(-\omega)} + \frac{\bar{a}^2 g^*(-\omega)}{h^*(-\omega)f^*(-\omega)f(\omega)} +\\
    \frac{\bar{a}^{*^2} g(\omega)}{h(\omega)f(\omega)f^*(-\omega)}) + \frac{i \hbar G \sqrt{2\kappa}  \hspace{2pt} \bar{b}^*}{\omega}
\end{split}
\end{equation}

\noindent and

\begin{equation}
\begin{split}
    \chi_{\dot{\phi}}(\omega) = i \hbar G \Big( 1+\frac{iG_B \vert \bar{b} \vert^2}{\omega}\Big) \Big( \frac{\vert a \vert^2 }{f(\omega)h(\omega)}+ \\
    \frac{\vert a \vert^2 }{f^*(-\omega)h^*(-\omega)} + \frac{\bar{a}^2 g^*(-\omega)}{h^*(-\omega)f^*(-\omega)f(\omega)} - \\
    \frac{\bar{a}^{*^2} g(\omega)}{h(\omega)f(\omega)f^*(-\omega)}\Big).
\end{split}
\end{equation}

The number of phonons can then be calculated as

\begin{equation}
    n_f=\frac{m\Omega_m^'}{\hbar}\int_{-\infty}^{\infty}S_{xx}(\omega)\frac{d\omega}{2\pi}
\end{equation}

\noindent where $\Omega_m^'$ is the shifted mechanical resonance frequency. Instead of integrating $S_{xx}(\omega)$ we consider the high mechanical quality factor approximation. In case the quality factor is large, the mechanical oscillator samples the various external noises only near the mechanical resonance frequency. Therefore, we integrate $|\chi_{eff}(\omega)|^2$ analytically as a Lorentzian function and evaluate the noise spectral densities at $\pm \Omega_m^'$. To ensure that this approximation is valid, we only accept results for which the final effective mechanical quality factor is more than 100.

 \vspace{1cm}

\bibliographystyle{nicebib}
\bibliography{References}

\end{document}